\documentclass[conference]{IEEEtran}

\usepackage[T1]{fontenc}
\usepackage{mathtools}
\usepackage[inline]{enumitem}

\usepackage{mathtools}

\usepackage[nolist]{acronym}

\usepackage{amsmath}  
\usepackage{algpseudocode} 
\usepackage{algorithmicx} 
\usepackage{float} 
\usepackage{caption} 

\usepackage{graphicx}
\usepackage[tight,footnotesize]{subfigure}
\usepackage{epstopdf}
\graphicspath{{Figures/}}

\usepackage{url}
\usepackage{multirow}
\usepackage{balance}
\usepackage[numbers,sort&compress]{natbib}
\usepackage{amsmath}
\usepackage{amssymb}
\usepackage{bm}
\usepackage{lipsum}
\usepackage{color}
\usepackage[table,xcdraw]{xcolor}
\usepackage{threeparttable}

\usepackage{physics} 

\usepackage{multirow}
\usepackage{authblk}

\usepackage[linesnumbered,ruled,vlined]{algorithm2e}

\SetCommentSty{mycommfont}

\SetKwInput{KwInput}{Input}                
\SetKwInput{KwOutput}{Output}              

\renewcommand{\baselinestretch}{.980}

\usepackage{lscape}
\usepackage{longtable}

\begin{document}

\title{Energy Saving in 6G O-RAN Using DQN-based
xApp}

\author{Qiao Wang}
\author{Swarna Chetty}
\author{Ahmed Al-Tahmeesschi}
\author{Xuanyu Liang}
\author{Yi Chu}
\author{Hamed Ahmadi}

\affil{School of Physics Engineering and Technology, University of York, United Kingdom}

\IEEEoverridecommandlockouts
\maketitle
\IEEEpubidadjcol

\begin{abstract}
Open \acf{RAN} is a transformative paradigm that supports openness, interoperability, and intelligence, with the O-RAN architecture being the most recognized framework in academia and industry. In the context of Open \ac{RAN}, the importance of \acf{ES} is heightened, especially with the current direction of network densification in \ac{6G}.
    Traditional energy-saving methods in \ac{RAN} struggle with the increasing dynamics of the network. 
    This paper proposes using \ac{RL}, a subset of \acf{ML}, to improve \ac{ES}. We present a novel deep \ac{RL} method for \ac{ES} in 6G O-RAN, implemented as xApp (\ac{ES}-xApp). We developed two \ac{DQN}-based \ac{ES}-xApps. \ac{ES}-xApp-1 uses RSS and \ac{UE} geolocations, while \ac{ES}-xApp-2 uses only RSS. The proposed models significantly outperformed heuristic and baseline xApps, especially with over 20 \acp{UE}. With 50 \acp{UE}, 50\% of \acp{RC} were switched off, compared to 17\% with the heuristic algorithm. We have observed that more informative inputs may lead to more stable training and results. This paper highlights the necessity of energy conservation in wireless networks and offers practical strategies and evidence for future research and industry practices.

\end{abstract}

\begin{IEEEkeywords}
O-RAN, 6G, Energy Saving, Deep Q-netowrk, Deep Reinforcement learning, Machine Learning, Reinforcement Learning
\end{IEEEkeywords}
\IEEEpeerreviewmaketitle


\section{Introduction}
Throughout the generations of mobile networks (currently \ac{5G}), the \acf{RAN} has been the imperative component. Thanks to the advancement in \ac{ICT}, the architecture of \ac{RAN} itself has evolved significantly in the past decades, from \ac{D-RAN} in the early generation to \ac{vRAN} in \ac{5G}. On top of \ac{vRAN}, the recent paradigm of \ac{RAN}, Open \ac{RAN} \cite{10024837}, further transforms the \ac{RAN} ecosystem. It aims to transform conventional RAN into an open, programmable, virtualized, interoperable, and intelligent architecture. The best-known architecture of Open \ac{RAN} is O-RAN which was proposed by the O-RAN Alliance. Key features of the O-RAN architecture include the standardized open interfaces to achieve interoperability and the novel \acf{RIC} \cite{oranwg1} component to enable \ac{AI} and \acf{ML} \cite{10552840} . O-RAN enables mobile operators to deploy \ac{ML} models as containerized xApp/rApp in \ac{RIC} to achieve automation and intelligence in various use cases. 

With the deployment of \ac{5G} underway globally, the research community has shifted its focus towards anticipating the next technological frontier. From the early works on \ac{6G} \cite{viswanathan2020communications} it can be realized that \ac{6G}, similar to \ac{5G}, will be driven by new applications and verticals such as autonomous systems and \ac{DT}  \cite{Ahmadi21DT}. 
On top of these, the pervasive use of \ac{AI} and \ac{ML} in \ac{6G} is what all existing works and researchers agree on, making \ac{6G} the most intelligent generation \cite{AIgreen21}. Given the fact that \ac{AI} and \ac{ML} will be an integrated component of both \ac{6G} and O-RAN, 6G RAN, powered by O-RAN, is believed to be the dominant \ac{RAN} architecture of future mobile networks.

However the \ac{RAN} evolves, the energy consumption of the network has always been a major concern. According to \cite{feng2012survey} in 2012, mobile operators were already reported as one of the top energy consumers. Moreover, the mass deployment of ultra-dense small-scale \acf{BS}, also known as network densification, is the major trend of \ac{5G} and future \ac{6G}. Despite the improvement in network capacity, coverage, data rate, and network latency brought by network densification, it inevitably significantly increases the total energy consumption of wireless systems. Hence, effective actions on energy saving are urgent. This is not only a crucial aspect of the operational cost of mobile operators but also important to environmental sustainability. An efficient and widely adopted energy-saving approach is to dynamically turn off underutilized \ac{BS}s (or \ac{RU}s)\footnote{We use BS and RU interchangeably in this paper.} while preserving the \ac{QoS} of mobile users, i.e., \ac{BS} ON-OFF switching \cite{8014292}. For example, some \ac{RU}s that may have few \ac{UE}s during certain time of the day can be considered to switch off or set to sleep mode and the associated \ac{UE}s can be steered to neighbouring \ac{RU}s if applicable. Another example is when a \ac{RU} is serving as a capacity layer of a macro \ac{BS} for \ac{eMBB} UEs but only has voice UEs connected. In such a case, the \ac{RU} can be switched off and hand over voice UEs to the macro \ac{BS}. In a nutshell, this approach requires decision-making strategies based on the network information.

In nature, \ac{BS} ON-OFF scheduling mechanism is combinational and is generally NP-hard \cite{8014292}, especially jointly considered with multiple performance metrics resulting in a mixed integer problem. Approaches to this problem in the literature were step-by-step greedy algorithms and iteratively solving decomposed subproblems \cite{6489498}. However, these hard-coded algorithms may not reach a global optimal solution either due to inadequate edge cases considered or inappropriate subproblems. On the other hand, this problem can be seen as a Markov Decision Problem (MDP) where the decisions or actions are BS ON-OFF. Therefore, \acf{RL} can be a potential solution. More precisely, given the discrete action space and model-free feature, value-based \ac{RL} method such as Q-learning \cite{sutton2018} is a feasible approach. Nevertheless, tabular Q-learning has the limitation in dealing with large state space which can be massive in mobile networks. To overcome this limitation, deep Q-learning was proposed that makes use of deep learning models (neural networks) as a Q-function approximator and \ac{DQN} \cite{mnih2015human} is a representative model for deep Q-learning. By learning the action values based on the states of the \ac{RAN} (e.g., user distribution, user association, user throughput, etc), a well-trained model can output the optimal action to switch on or off an \ac{RU}, given a particular network state.

The purpose of this paper is to present a novel \ac{DQN}-based approach to dynamic \ac{RU} ON-OFF switching to improve energy saving in O-RAN. Our contributions can be outlined as follows:
\begin{itemize}
    \item We design a novel \ac{DQN}-based algorithm for energy saving in O-RAN by optimizing the strategy of switching off \acp{RC} in 6G O-RAN network. The \ac{DQN} model is further developed as two different \ac{ES}-xApps which take different types of network information as input state. 
    \item We conduct benchmarking with these two xApps against a heuristic xApp \cite{liang2024enhancing}, a rule-based one, and a baseline xApp. Significant advantage of our proposed method is demonstrated. We also demonstrate the robustness of the \ac{DQN} model with more informative information over its counterpart.   
\end{itemize}

The rest of the paper is structure as follows. In Section II, we provide an overview of the background of O-RAN and energy saving. We describe our system model and formulate an optimization problem for energy saving in Section III. In Section IV, we introduce the technical details of the proposed \ac{DQN} model and the algorithm for training. We discuss the simulation environment and analyse the performance in Section V. Section VI concludes the paper. 


\section{Background Overview}

\subsection{O-RAN overview}

Formed in February 2018 by merging xRAN Forum and \ac{C-RAN} Alliance, O-RAN Alliance has been leading the standardization of Open RAN. The \ac{3GPP} 5G RAN is supporting O-RAN in many ways such as the split option 7.2 and major \ac{MNO}s such as Vodafone are already starting to evolve their RANs to become O-RAN compatible.  
The main objective of O-RAN is to steer the \ac{RAN} industry toward openness, intelligence, virtualization, and complete interoperability. 

The two main pillars of O-RAN are \textit{openness} and \textit{intelligence}. The aim of openness is to eliminate vendor lock-in and proprietary equipment through open and standardized interfaces. This is achieved in O-RAN by newly defined open interfaces such as A1, O1, and E2. Openness is a prerequisite for multi-vendor and interoperable RAN, which gives the MNOs the most flexibility to deploy their RAN using network components from different vendors and hence accelerating the delivery of new features and services to mobile users. 

The embedded intelligence in O-RAN brings the automation in deployment, optimization, and operation of mobile networks. This is of great significance in 6G due to the increasingly complex system. O-RAN introduces RAN intelligence through the RAN Intelligent Controller (RIC). Two innovative RICs are designed by O-RAN: near-RealTime (near-RT) RIC and non-RealTime (non-RT) RIC. The \ac{RIC}s play an important role in enabling the intelligence in O-RAN. They are the key components in the AI/ML workflow to deliver intelligent management, control, and optimization decisions to the RAN nodes.

\subsection{Energy saving in O-RAN}
The past few decades have witnessed rapid growth and expansion in the telecommunications industry. However, the corresponding energy consumption has also reached unprecedented levels, especially as wireless networks have become denser \cite{9678321}. Therefore, \ac{ES} has become an important consideration in RAN design and operation from the perspective of mobile network operators. More importantly, \ac{ES} not only has enormous ecological benefits and represents a socially responsible response to climate change, but also has significant economic benefits \cite{feng2012survey}.

In wireless networks, the radio access part i.e., \ac{RAN} consumes more than 50\% of energy, where \ac{BS}s are the main contributor to this \cite{feng2012survey}. 
In practice, the mechanism of dynamically switching BSs has become a widely used strategy for energy saving. More precisely, the cells or carriers of a BS in a low load state can be turned off and back to operation whenever needed. However, this approach will pose significant challenges in maintaining the \ac{QoS} of the associate users. Thus, how to avoid degradation of network \ac{QoS} while taking these operations is an important topic. 

In O-RAN, \ac{ES} is still an important use case as O-RUs account for a large part of the energy consumption \cite{oranwg1-ee}. Therefore, similar strategies described above can also be applied to O-RUs. A very recent work \cite{liang2024enhancing} used a heuristic approach for this purpose with a commercial O-RAN \ac{RIC} tester. In addition, the innovative RICs in O-RAN can integrate \ac{ML} capabilities, making dynamic switching in O-RUs more intelligent than ever before. This is made possible by the specified open interfaces through which the RICs can access massive amounts of network data and use it to enhance ML models. Whether an O-RU can be switched off depends on many factors in the network such as user association, user distribution, signal strength, etc. Given the complexity of the modern RAN, this decision is difficult to implement manually. The advancement in machine learning makes this challenge addressable. 


\section{System Model and problem formulation}
\subsection{System model}
In this paper, we consider a small cell network \(\mathcal{G}\) in the O-RAN architecture. There are \(N\) O-RUs \footnote{The prefix 'O-' means O-RAN compatible components} in a set \(\mathcal{N}=\left \{ n_{1}, n_{2},..., n_{N}\right \}\) and they are all connected to a unique O-DU and a single O-CU. The O-RUs are positioned at fixed locations in \(\mathcal{G}\) with 100 meters inter-site separation, forming a dense environment. In addition, we consider these O-RUs as neutral hosts, each of which accommodates two \acp{RC} operating on distinct frequency bands, N77 and N78, respectively. Thus, there are \(M = 2N\) \ac{RC}s for \(N\) O-RUs and they are denoted with set \(\mathcal{M}=\left \{ m_{1}, m_{2},..., m_{M}\right \}\). \(K\) stationary \ac{UE}s in the set \(\mathcal{U} = \left \{u_{1}, u_{2}, ..., u_{K}\right \}\) are randomly distributed in the network and associated with a corresponding \ac{RC} \(m_{i} \in \mathcal{M}\). The free space path loss is considered and defined as:
\vspace{-2mm}
\begin{equation}
\text{FSPL} = 20 \log_{10}(d) + 20 \log_{10}(f) + 20 \log_{10}\left(\frac{4 \pi}{c}\right),
\end{equation}
where $d$ is the distance between the UE and serving \ac{RC}. $f$ is the transmission frequency in Hz (N77 or N78 in our case) and $c$ is the speed of light.

As introduced earlier, BS ON-OFF is an effective approach to energy saving in mobile networks. This is also the approach adopted in this paper. In our case, \acp{RC} are the associated cells for \ac{UE}s so in the following we refer our approach as \textit{\ac{RC} ON-OFF}. The core idea behind \ac{RC} ON-OFF is that the energy consumption of the network can be minimized if underutilized \acp{RC} can be switched off. However, this approach is likely to violate \ac{QoS} of the network and \ac{UE}s. For instance, an \ac{RC} has limited bandwidth which results in a limited capacity for serving \acp{UE}. If this capacity is breached, some \acp{UE} may end up with no available resources for their service such as data traffic. Therefore, it is of great importance to consider \ac{QoS} of the network and UEs as constraints. 

\subsection{Problem formulation}
We define the objective of our approach to switching \acp{RC} ON-OFF to minimize energy consumption through turning off as many \acp{RC} as possible without compromising network performance and \ac{QoS}. In other words, we are maximizing the number of underutilized \ac{RC}s that are switched off to achieve energy saving. For readability, we use \(\mathcal{Z}\) to denote the number of \ac{RC}s that are switched off.

For this optimization problem, we consider the following key constraints. The first is Received Signal Strength (RSS), which represents the strength of the received signal of UEs and affects the data rate. We use \(R_{k, m}\) to represent the RSS between \ac{UE} \(k \in \mathcal{U}\) and \ac{RC} \(m \in \mathcal{M}\). We assume that there exists a minimum requirement \(R_{min}\) for \(R_{k, m}\) for a given valid connection between \ac{UE} \(k\) and \ac{RC} \(m\). Thus, the following equation should be met:
\begin{equation}
R_{k, m} \geq R_{min},
\label{eq:rss:threshold}
\end{equation}
where \(R_{k, m}\) denotes the RSS for a valid link between \ac{UE} \(k\) and \ac{RC} \(m\). Additionally, we assume that each UE can only be served by a single \ac{RC}. Therefore, (\ref{eq:rss:threshold}) can be generalized as follows:
\begin{equation}
    \sum_{m =1}^{\mathcal{M}} \alpha_{k,m} R _{k,m}\geq R_{min}, \quad \forall k \in \mathcal{U}
\end{equation}
where \(\alpha_{k,m} \in \{0, 1\}\) is a binary indicator indicating whether a link between \ac{UE} \(k\) and \ac{RC} \(m\) exist

The second constraint is the number of \acp{UE} conected to an \ac{RC}. The \ac{RC}'s capacity \(C_{max}\) represents the maximum number of connections (i.e., the maximum number of UEs that an \ac{RC} can handle). We assume that all \acp{RC} have the same maximum capacity \(C_{max}\), so the following constraint should be satisfied:
\begin{equation}
C_m \leq C_{max}, \quad \forall m \in \mathcal{M}
\label{eq:cap:threshold}
\end{equation}

Consequently, the problem that our RC ON-OFF approach aims to address can be formulated as follows:
\begin{equation}
\begin{aligned}
\mathcal{P}: \quad \max \quad &  \mathcal{Z} \\
\text{s.t.} \quad & C_m \leq C_{max}, \quad \forall m \in \mathcal{M}, \\
                    & \sum_{m =1}^{\mathcal{M}} \alpha_{k,m} R_{k,m}\geq R_{min}, \quad \forall k \in \mathcal{U},\\
                    & \sum_{m =1}^{\mathcal{M}}\alpha_{k,m}\leq 1 \quad \forall k \in \mathcal{U}
\end{aligned}
\label{eq:op}
\vspace{-4mm}
\end{equation}

\begin{figure}[!t]
	\centering
	\includegraphics[clip, trim=0.0cm 0.cm 0.0cm 0cm, width=1\columnwidth]{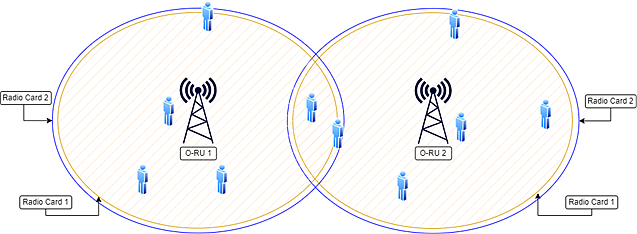}
	\caption{Network deployment with each \ac{RU} having two \acp{RC}.}
	\label{fig:ES-scenario}
\vspace{-4mm}
\end{figure}

\section{Deep reinforcement learning solution}
To solve the above problem, we propose to use the \ac{RL} technique to solve it because the \ac{RC} ON-OFF is essentially a decision-making process and can be modeled as a Markov Decision Process (MDP). More specifically, \ac{RC} ON-OFF is the action to take and this forms a discrete action space. For such a finite discrete action space, we can use \(Q\)-learning \cite{sutton2018} to solve it. However, traditional \(Q\)-learning use \(Q\)-table to update \(Q\)-values of state-action pairs, which can be infeasible in wireless networks where we are facing massive state space. Therefore, deep \(Q\)-learning was proposed to use neural networks as function approximator and output \(Q\)-values for each action, and this is referred to as \acf{DQN} \cite{mnih2015human}. 
In the following, we will introduce our \ac{DQN}-based algorithm to address the optimization problem in (\ref{eq:op}).
We will start with a brief basics of deep \(Q\)-learning. 

\subsection{Basics of Deep Q-learning}
The feature that differentiates \ac{RL} from supervised or unsupervised learning is the presence of an \textit{environment} and an \textit{agent}. During the learning process, the RL agent interacts with the environment without making any assumptions about its structure or properties. At discrete time intervals, the agent takes an action based on the current state of the environment. The environment then transitions to the next state as a result of the action and the previous state. The outcome of the action is a reward, which helps to optimize the current policy. The objective of the agent is to obtain its optimal policy \(\pi^* = \mathop{\arg\max}_{\pi} \mathbb{E}\left[\sum_{t=0}^{\infty}\gamma^t r_t \mid \pi\right]\) to maximize its return, i.e., the expected cumulative reward, through interactions with the environment.

$Q$-learning is a classic model-free, value-based \ac{RL} algorithm which learns the optimal action-value \(Q_*(s, a)=\max_\pi Q_\pi(s,a)\) and derives the optimal policy \(\pi^*\) from the optimal values by selecting the highest-valued action in each state. However, most practical problems are too large to learn all action values in all states separately. Thus, \ac{DQN} uses a multi-layered neural network to output a vector of action values \(Q(s, \cdot; \theta)\) for a given state \(s\), where \(\theta\) are the parameters of the network. One important ingredient of \ac{DQN} is the use of a target network with parameters \(\theta'\). It is the same as the policy network with \(\theta\), except that its parameters \(\theta'\) are copied every \(\tau\) steps from the policy network and kept fixed for the remaining steps to result in a stable target
  \(Y_{t}^{DQN} = R_{t+1} + \gamma max_a Q(s_{t+1}, a; \theta')\)
Thus, the parameters \(\theta\) of the policy network is updated by minimizing the loss function 
\begin{equation}
\label{eq:loss}
    L(\theta) = (Y_{t}^{DQN} - Q(s_t, a_t; \theta))^2
\end{equation}

The other important ingredient is experience replay where the stored transitions are sampled uniformly to update the network. 

\subsection{DQN-based energy saving model}
In this subsection, we discuss the state space, action space, and reward function of the proposed DQN-based energy saving solution. The network model consists of \(M\) \ac{RC}s and \(K\) \ac{UE}s, which is considered as the environment, and the xApp that hosts our DQN model is serving as an agent. Based on this setting, we can define the state space, action space, and reward function as follows.

\subsubsection{State space}
The state of the environment (i.e., O-RAN network) contains essential information that should be fed into the \ac{DQN}. In our proposed \ac{DQN} framework, we consider a user-centric state. More specifically, the state vector observed from the network is constructed from \ac{UE}-related information. We use \(\mathbf{H}^{(t)}\) to represent the RSS matrix of UEs at timestamp \(t\), resulting in a \(M \times K\) matrix as:
\begin{equation}
\mathbf{H}^{(t)} = \begin{bmatrix}
h_{1,1}^{(t)} & \cdots & h_{1,K}^{(t)} \\
\vdots & \ddots & \vdots \\
h_{M,1}^{(t)} & \cdots & h_{M,K}^{(t)}
\end{bmatrix}
\label{eq:rss}
\end{equation}
Each column of \(\mathbf{H}^{(t)}\) represents the RSSs of the \(M\) \ac{RC}s received by a UE \(k \in K\) sorted in descending order. Since the UEs are all static, the top RSS is usually from its serving \ac{RC}. 

In addition to the RSS information, UEs' geolocations are also considered. We denote the \(K\) UEs' x- and y-coordinates in the vector of tuples \(\mathbf{G}^{(t)}=\left[ (x_1, y_1), (x_2, y_2), \cdots, (x_K, y_K) \right]\). Our \ac{DQN} model takes the input state as a vector which consists of \(\mathbf{H}^{(t)}\) and \(\mathbf{G}^{(t)}\): 
\begin{equation}
    \mathbf{s}^{(t)} = \left[ \mathbf{G}^{(t)}(1) \mathbf{H}^{(t)}(:, 1),\cdots, \mathbf{G}^{(t)}(K) \mathbf{H}^{(t)}(:, K)   \right]^\top,
\label{eq:rss+geo}
\end{equation}

where \(\mathbf{G}^{(t)}(k) \) represents the coordinates of UE \(k\in K\) and \(\mathbf{H}^{(t)}(:,k)\) for RSS information received by UE \(k\).
\subsubsection{Action space}
The size of the action space is determined by the number of \acp{RC}. Since each \ac{RC} has two states - ON or OFF, the resulting action space is of size $\mathcal{A} = $\(2 \times M\):
\begin{equation}
    \mathbf{A} = [\text{on}_{1},\text{off}_{1}, \cdots, \text{on}_{M},\text{off}_{M}  ]
\end{equation}
At each time step, an action \(a^{(t)} \in \mathbf{A}\) will be selected. 

\subsubsection{Reward function}
Recall that the problem we are solving is to maximize the number of underutilized \acp{RC} that are switched off to achieve energy saving while not violating \ac{QoS}. Therefore, it is important that the reward \(r_t\) at each time step reflects the optimization objective and the constraints in (\ref{eq:op}). We define our reward \(r_t\) as following:

\begin{equation}
    r_t = r_\text{RC-OFF} + r_\text{RSS-Breach} + r_\text{Capacity-Breach}
\end{equation}

Specifically, \(r_\text{RC-OFF}\) is calculated as:
\begin{equation}
r_\text{RC-OFF} = -5 \times \text{ratio of OFF \acp{RC}} + 1 \times \text{ratio of ON \acp{RC}}
\end{equation}

\(r_\text{RSS-Breach}\) is calculated based on whether the RSS of \acp{UE}'s serving \ac{RC} breaches the \ac{QoS} threshold of $R_{min} = -95$ dBm, as in (\ref{eq:rss:threshold}):

\begin{equation}
r_\text{RSS-Breach} = 
\begin{cases}
5 & \sum_{m =1}^{\mathcal{M}} \alpha_{k,m} R_{k,m}\geq R_{min}, \quad \forall k \in \mathcal{U}\\
-20 & \text{otherwise}
\end{cases}
\end{equation}

\(r_\text{Capacity-Breach}\) is computed based on the check if any \acp{RC}'s number of connections is larger than $C_{max}$ ($C_{max}$ is set to 10 in this work), as in (\ref{eq:cap:threshold}):
\begin{equation}
r_\text{Capacity-Breach} = 
\begin{cases}
1 & \text{for each non-violated \ac{RC}}\\
-1 & \text{for each oversubscribed \ac{RC}}
\end{cases}
\end{equation}

\subsection{Proposed DQN training and application}
The training process for the proposed DQN algorithm for \acp{RC} reconfiguration is outlined in Algorithm \ref{alg_DQN_training}. The algorithm begins with initializing the parameters (Lines 1–3), followed by running \(P\) episodes. Each episode consists of \(T\) steps and begins by resetting the state (Line 5). For action selection (Lines 7 and 8), the \(\epsilon\)-greedy strategy is used. After an action \(a_t\) is executed, a reward \(r_t\) is obtained, and the system transitions to a new state \(s_{t+1}\). The transition tuple \(\{s_t, a_t, r_t, s_{t+1}\}\) is stored in the experience replay memory \(D\) (Line 10). Once \(D\) exceeds the mini-batch size, training of the network begins (Lines 11–15). Random mini-batches of transitions are sampled from \(D\) to train the DQN with historical data. The target network is used to compute the target value \(y_t\) (Line 13), which is utilized to assess the actions chosen by the main Q-network. The loss function defined in (\ref{eq:loss}) is employed to update the main Q-network parameters \(\theta\) (Line 14). The target network parameters \(\theta'\) are updated every \(\tau\) training steps (Line 15).

\begin{figure}[!t]
	\centering
	\includegraphics[clip, trim=0.0cm 5.cm 0.0cm 2cm, width=1\columnwidth]{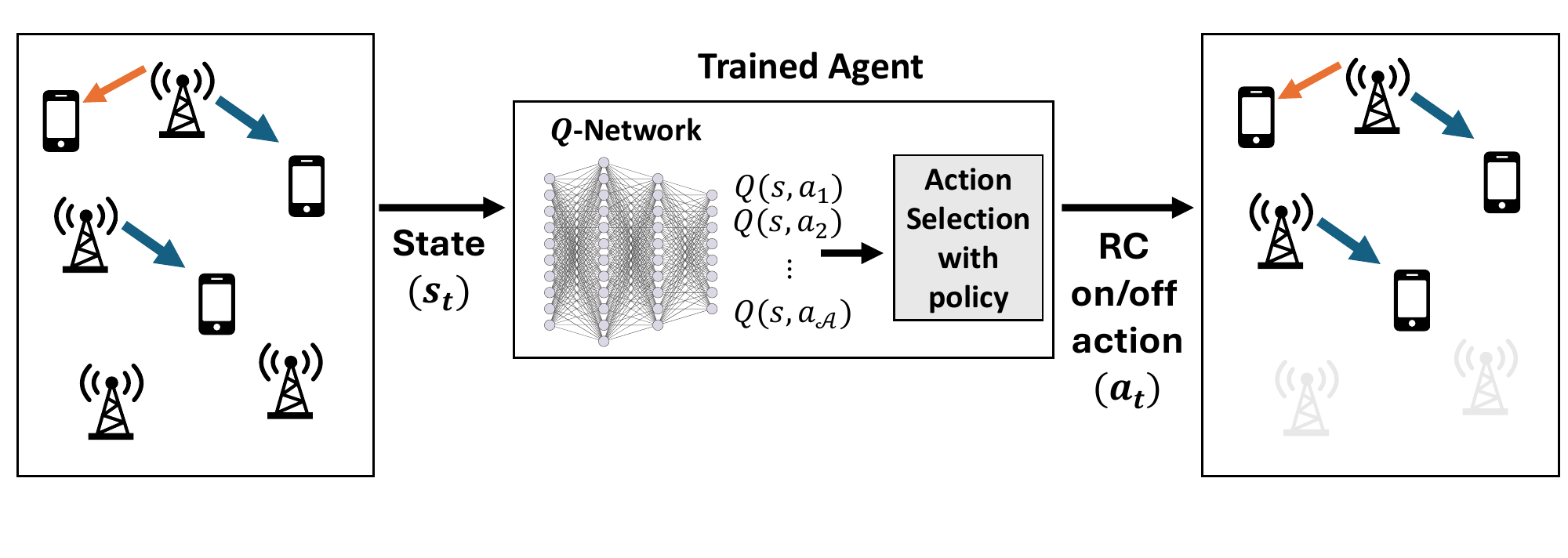}
	\caption{DRL based energy saving xApp.}
	\label{fig:flowgraph}
\end{figure}

Upon completion of the training, the proposed DQN is capable of determining the minimum number of active \acp{RC} necessary to support the \acp{UE} within the area of interest. During the inference phase as depicted in Figure \ref{fig:flowgraph}, the trained DQN monitors the environment state \(s_t\) at each step and selects an action that minimizes the number of active \acp{RC} while maintaining the quality of service for the \acp{UE}.

\begin{algorithm}[t]
\DontPrintSemicolon
  \KwInput{Vector representation of the states.}
  \KwOutput{O-RC configuration.}
  
Initialize the replay memory $D$ with a predefined maximum capacity \\
Initialize the Q-network weights $\theta$ \\
Initialize the target network with weights $\theta'$ equal to $\theta$ \\

  \For{episode = 1 to P} 
    {
        Initialize the environment and receive the initial state $s_t$ \\
        \For{step $t = 1$ to $T$} 
        {
            With probability $\epsilon$, choose a random action $a_t$ \\
            Otherwise, choose $a_t = \arg\max_{a} Q(s_t, a; \theta)$ \\
            Observe reward $r_t$ and the new state $s_{t+1}$ \\
            Store the transition $\{s_t, a_t, r_t, s_{t+1}\}$ in $D$ \\
            
            \If{memory is full}
            {
                Randomly sample a mini-batch of transitions $\{s_t, a_t, r_t, s_{t+1}\}$ from $D$ \\
                Set $y_t = \begin{cases}
                    r_t & \text{if } t = T\\ 
                    r_t + \gamma \max_{a} Q(s_{t+1}, a; \theta') & \text{if } t < T
                \end{cases}$ \;
                Update the weights $\theta$ of the main Q-network by minimizing the loss function \\
                Update $\theta'$ to $\theta$ every $\tau$ steps
            }
        }
    }
\caption{DRL-based solution for RC state switching (training phase)}
\label{alg_DQN_training}
\end{algorithm}

\section{Performance Analysis}
\subsection{Evaluatoin scenario}

\begin{table}[]
\begin{tabular}{|l|l|}
\hline
Parameters                        & Value                          \\ \hline
Number of UEs                     & {[}10, 20, 30, 50{]}   \\ \hline
Number of O-RUs                     & 6                              \\ \hline
Number of \ac{RC}s per O-RU       & 2                              \\ \hline
Channel model                     & Free space path loss           \\ \hline
Transmission power per \ac{RC} & 30 dBm                         \\ \hline
Frequency                         & N77 and N78                         \\ \hline
Handover                          & Received signal strength based \\ \hline
Similation area                   & 500 m x 500 m                    \\ \hline
\end{tabular}
\caption{Network simulation parameters}
\label{tab1}
\end{table}

To evaluate the performance of the proposed \ac{DQN} solution, we conducted the experiments in a simulated O-RAN environment. Specifically, we developed an O-RAN simulator to simulate the \ac{RAN} environment in Python 3.10 and the relevant settings of the environment can be found in Table \ref{tab1}. In addition, we developed the proposed \ac{DQN} solution as two \ac{ES}-xApps. \ac{ES}-xApp-1 takes both RSS and UEs' geolocations as state i.e., \(\mathbf{s}\) in (\ref{eq:rss+geo}), while the \ac{ES}-xApp-2 only takes RSS i.e., \(\mathbf{H}\) in (\ref{eq:rss}) as \ac{DQN} state. 

We performed the training of the \ac{ES}-xApps with Pytorch framework for 30000 episodes with 100 steps per episode. AdamW optimizer with a learning rate of \(L_r = 10^{-4}\) was used. The discount parameter \(\gamma\) is set to 0.99, the mini-batch size is 64, the memory buffer is set to 300000, and the target network parameters \(\theta'\) are updated every 50 steps. Moreover, we used epsilon decay for the exploration-exploitation strategy, where \(\epsilon\) starts from 0.9 reducing to 0.05. At the inference stage, the \ac{ES}-xApps are able to collect \ac{KPM}s from the O-RAN environment to construct \ac{DQN} states and send action to the environment to control the \ac{RC}s.

\subsection{Simulation results}
To evaluate the effectiveness of the \ac{DQN}-based ES-xApps in the inference stage, we conducted a comparative analysis against a baseline xApp for toggling O-RU \acp{RC} on and off. This baseline xApp operates by assessing whether any UEs are connected to an \ac{RC}. If no UEs are linked, the algorithm proceeds to switch off that \ac{RC}. We also compared them with a heuristic algorithm recently published \cite{liang2024enhancing}.

\begin{figure}[!htbp]
        \vspace{-5mm}
    \centerline{\includegraphics[clip, trim=0.0cm 0cm 0.0cm 0cm,width = 1.05\columnwidth]
{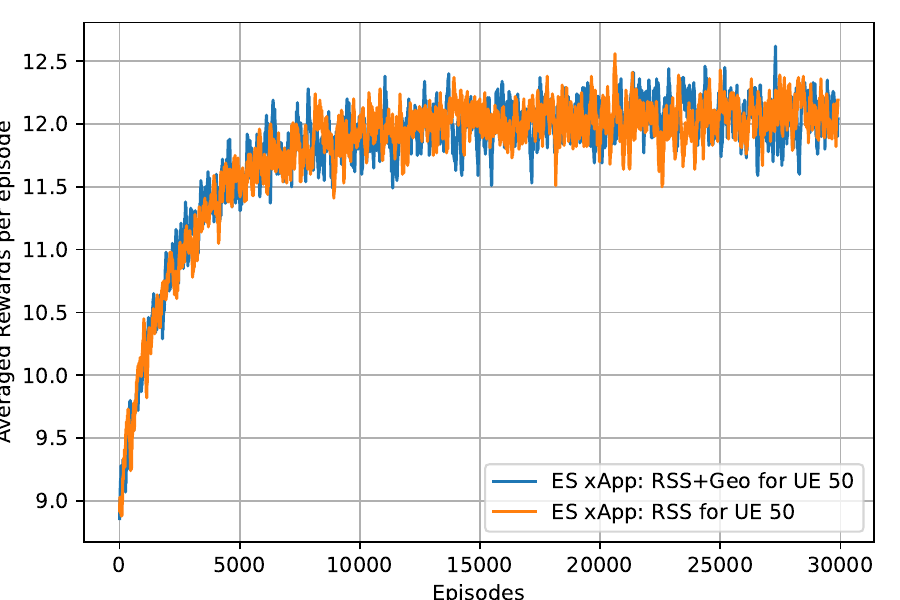}}
    \caption{Moving average reward of two ES-xApps with a sliding window of 50 across episodes for 50 UEs scenario}
    \label{fig3}
    \end{figure}

\begin{figure}[!htbp]
    \centerline{\includegraphics[clip, trim=0.0cm 0cm 0.0cm 0cm,width = 1.05\columnwidth]{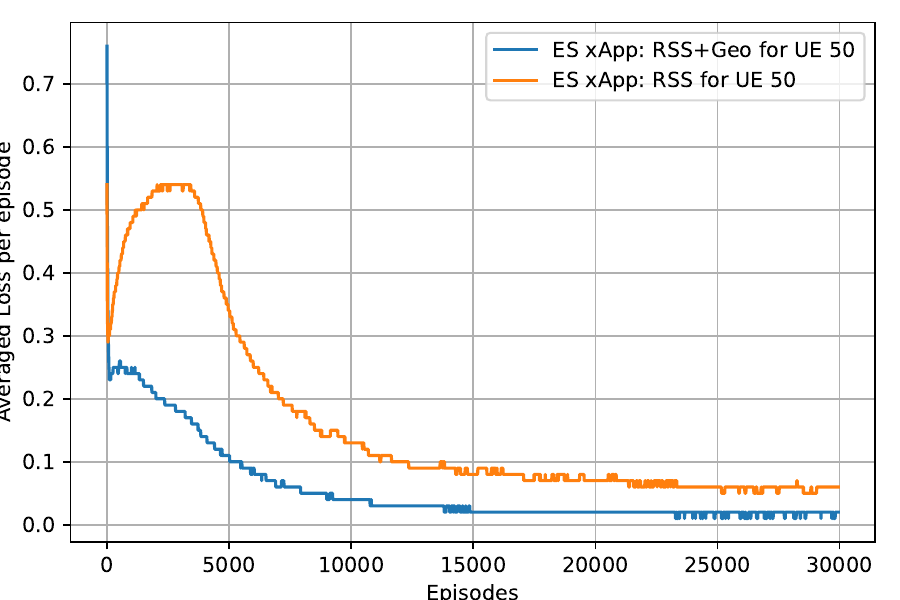}}
    \caption{Moving average loss of two ES-xApps with a sliding window of 50 across episodes for 50 UEs scenario}
    \label{fig4}

    \end{figure}

Fig. \ref{fig3} and Fig. \ref{fig4} illustrate the moving average for the reward and loss of the two ES-xApps, respectively, where we selected the configuration of 50 \acp{UE} as a representative example. While there is no major difference in the average reward, the ES-xApp-1 that inputs the additional UEs' geolocation information showed a faster and more stable convergence. Similar trends have been observed in other \ac{UE} configurations, which means that a multi-dimensional input state may help with the training process. 

\begin{figure}[!htbp]
\centerline{\includegraphics[clip, trim=0.0cm 0cm 0.0cm 0cm,width = 1.05\columnwidth]{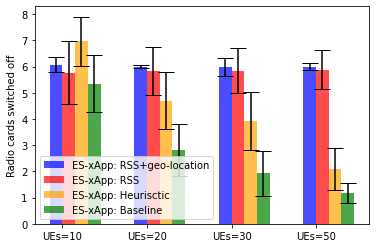}}
    \caption{Performance benchmarking between two ES-xApps, heuristic xApp, and baseline xApps}
    \label{fig5}
    \vspace{-8mm}
    \end{figure}

Fig. \ref{fig5} shows the average number of \acp{RC} that are switched off for the four algorithms with standard deviation based on 50 simulations. Both of our \ac{DQN}-based xApps outperformed the heuristic and the baseline increasingly significantly for UE configurations 20, 30, and 50. For 10 \acp{UE}, even though the heuristic algorithm turned off more \acp{RC} on average, our \ac{ES}-xApp-1 shows a much smaller standard deviation, which indicates a less variability or dispersion from the average value.
As depicted by the error bars, an additional observation is that including \acp{UE}' geolocation in the input state of \ac{DQN} also leads to more stability and less variation compared to its counterpart.

    \vspace{-8mm}

\section{Conclusions}
\acf{ES} in wireless networks stands as a pivotal concern, impacting both economic sustainability and environmental conservation. To address this challenge, we formulated an optimization problem and used \acf{DRL} algorithm to solve it. Concretely, we developed two novel \ac{DQN}-based \ac{ES}-xApps in the 6G O-RAN architecture. \ac{ES}-xApp-1 takes both RSS and UEs' geolocations as state, while the \ac{ES}-xApp-2 only takes RSS. Simulation results showed that the trained models outperformed the heuristic and baseline xApps especially when the number of \acp{UE} is over 20. When reaching 50 \acp{UE}, 50\% of the \acf{RC}s are switched off to save energy, while this is only 17\% using the heuristic algorithm. Another important observation is that more useful information such as UEs' geolocations in the input state may result in more stable training and less variation in the results.
This paper not only provides invaluable insights into the imperative of energy saving in wireless networks but also offers actionable strategies and empirical evidence to guide future research and industry practices.

\section*{Acknowledgement}

This work was supported in part by the Engineering and Physical Sciences Research Council United Kingdom (EPSRC), Impact
Acceleration Accounts (IAA) (Green Secure and Privacy Aware Wireless Networks for Sustainable Future Connected and Autonomous
Systems) under Grant EP/X525856/1 and Department of Science, Innovation and Technology, United Kingdom, under Grants Yorkshire
Open-RAN (YO-RAN) TS/X013758/1 and RIC Enabled (CF-)mMIMO for HDD (REACH) TS/Y008952/1.

\begin{acronym} 
\acro{5G}{the fifth generation of mobile networks}
\acro{6G}{sixth generation of mobile networks}
\acro{3GPP}{3rd Generation Partnership Project}
\acro{ACO}{Ant Colony Optimization}
\acro{AI}{Artificial Intelligence}
\acro{AR}{Augmented Reality}
\acro{BB}{Base Band}
\acro{BBU}{Base Band Unit}
\acro{BCI}{Brain Computer Interface}
\acro{BER}{Bit Error Rate}
\acro{BS}{Base Station}
\acro{BW}{bandwidth}
\acro{C-RAN}{Cloud Radio Access Networks}
\acro{CAPEX}{Capital Expenditure}
\acro{CoMP}{Coordinated Multipoint}
\acro{CPU}{Central Processing Unit}
\acro{CR}{Cognitive Radio}
\acro{CRN}{Cognitive Radio Network}
\acro{D2D}{Device-to-Device}
\acro{DA}{Digital Avatar}
\acro{DAC}{Digital-to-Analog Converter}
\acro{DAS}{Distributed Antenna Systems}
\acro{DBA}{Dynamic Bandwidth Allocation}
\acro{DC}{Duty Cycle}
\acro{DL}{Deep Learning}
\acro{DQN}{Deep Q-Network}
\acro{DRAM}{Dynamic Random Access Memory}
\acro{DRL}{Deep Reinforcement Learning}
\acro{DSA}{Dynamic Spectrum Access}
\acro{DT}{Digital Twin}
\acro{D-RAN}{Distributed Radio Access Network}
\acro{ES}{Energy Saving}
\acro{eMBB}{Enhanced Mobile Broadband}
\acro{FBMC}{Filterbank Multicarrier}
\acro{FEC}{Forward Error Correction}
\acro{FFR}{Fractional Frequency Reuse}
\acro{FSO}{Free Space Optics}
\acro{GA}{Genetic Algorithms}
\acro{GPU}{Graphic Processing Unit}
\acro{HAP}{High Altitude Platform}
\acro{HL}{Higher Layer}
\acro{HARQ}{Hybrid-Automatic Repeat Request}
\acro{IoT}{Internet of Things}
\acro{ICT}{Information and Communications Technology}
\acro{KPI}{Key Performance Indicator}
\acro{KPM}{Key Performance Measurement}
\acro{LAN}{Local Area Network}
\acro{LAP}{Low Altitude Platform}
\acro{LL}{Lower Layer}
\acro{LOS}{Line of Sight}
\acro{LTE}{
Long Term Evolution}
\acro{LTE-A}{Long Term Evolution Advanced}
\acro{MAC}{Medium Access Control}
\acro{MAP}{Medium Altitude Platform}
\acro{MIMO}{Multiple Input Multiple Output}
\acro{ML}{Machine Learning}
\acro{MME}{Mobility Management Entity}
\acro{mmWave}{millimeter Wave}
\acro{MNO}{Mobile Network Operator}
\acro{MR}{Mixed Reality}
\acro{NASA}{National Aeronautics and Space Administration}
\acro{NFP}{Network Flying Platform}
\acro{NFPs}{Network Flying Platforms}
\acro{NTNs}{Non-terrestrial networks}
\acro{NFV}{Network Function Virtualisation}
\acro{NN}{neural network}
\acro{OAM}{Orbital Angular Momentum}
\acro{OFDM}{Orthogonal Frequency Division Multiplexing}
\acro{OSA}{Opportunistic Spectrum Access}
\acro{PAM}{Pulse Amplitude Modulation}
\acro{PAPR}{Peak-to-Average Power Ratio}
\acro{PGW}{Packet Gateway}
\acro{PHY}{physical layer}
\acro{PSO}{Particle Swarm Optimization}
\acro{PT}{Physical Twin}
\acro{PU}{Primary User}
\acro{QAM}{Quadrature Amplitude Modulation}
\acro{QoE}{Quality of Experience}
\acro{QoS}{Quality of Service}
\acro{QPSK}{Quadrature Phase Shift Keying}
\acro{RF}{Radio Frequency}
\acro{RL}{Reinforcement Learning}
\acro{RN}{Remote Node}
\acro{RRH}{Remote Radio Head}
\acro{RU}{Radio Unit}
\acro{RC}{Radio Card}
\acro{RRC}{Radio Resource Control}
\acro{RRU}{Remote Radio Unit}
\acro{RAN}{Radio Access Network}
\acro{RIC}{RAN Intelligent Controller}
\acro{SU}{Secondary User}
\acro{SCBS}{Small Cell Base Station}
\acro{SDN}{Software Defined Network}
\acro{SMO}{Service Management and Orchestration}
\acro{SNR}{Signal-to-Noise Ratio}
\acro{SON}{Self-organising Network}
\acro{TDD}{Time Division Duplex}
\acro{TD-LTE}{Time Division LTE}
\acro{TDM}{Time Division Multiplexing}
\acro{TDMA}{Time Division Multiple Access}
\acro{UE}{User Equipment}
\acro{UAV}{Unmanned Aerial Vehicle}
\acro{USRP}{Universal Software Radio Platform}
\acro{VNF}{Virtual Network Function}
\acro{vRAN}{Virtualized Radio Access Network}
\acro{VR}{Virtual Reality}
\acro{XAI}{Explainable Artificial Intelligence}
\end{acronym}

\footnotesize
\bibliographystyle{IEEEtran}
\bibliography{References}

\end{document}